\documentclass{raa}                          

\usepackage{graphicx,times}             
\usepackage{natbib}
\usepackage{amssymb,amsmath}
\bibpunct{(}{)}{;}{a}{}{,}
\usepackage[pagebackref=true]{hyperref}

\begin{document}

  \title{Multiple blobs following a coronal mass ejection}

   \volnopage{Vol.0 (2026) No.0, 000--000}
   \setcounter{page}{1} 

   \author{Qing-Min Zhang
      \inst{1}
   \and  Jin-Han Guo
      \inst{2}
   \and Jie Chen
      \inst{3}
   \and Zong-Yi Li
      \inst{1,4}
  \and Yi-Wei Ni
      \inst{5}
  \and Yan-Jie Zhang
      \inst{1}
   }

   \institute{Purple Mountain Observatory, Chinese Academy of Sciences, Nanjing 210023, China; {\it zhangqm@pmo.ac.cn}\\
        \and
             State Key Laboratory of Solar Activity and Space Weather, School of Aerospace, Harbin Institute of Technology, Shenzhen 518055, China\\
        \and
             National Astronomical Observatories, Chinese Academy of Sciences, Beijing 100012, China\\
        \and
             School of Astronomy and Space Science, University of Science and Technology of China, Hefei 230026, China\\
        \and
             School of Astronomy and Space Science, Nanjing University, Nanjing 210023, China\\
\vs\no
   {\small Received 2026 month day; accepted 2026 month day}}

\abstract{In this paper, we study a partial filament eruption, 
which generated a C4.1 class two-ribbon flare and a wide coronal mass ejection (CME) propagating in the northeast direction on 2023 March 20. 
The long, sigmoidal filament in active region 13258 is composed of two parts: the eruptive part and remaining part. Both of them are supported by magnetic flux ropes.
The separation of two ribbons persist for $\sim$12 hr, 
with the separation speeds decreasing from 0.26$-$5.62 to 0.13$-$0.53 km\,s$^{-1}$ observed in H$\alpha$ and 304 {\AA}.
A thin, curved current sheet follows the CME and extends in the northeast direction. 
Eight and seven bright blobs are found to move outward along the current sheet observed by 
the Large Angle Spectroscopic Coronagraph (LASCO) on board the Solar and Heliospheric Observatory (SOHO) mission
and the COR2 coronagraph on board the ahead Solar TErrestrial RElations Observatory (STEREO), respectively.
In LASCO images, the apparent speeds, lifetime, maximal height, and width of the blobs are 
378$-$871 km\,s$^{-1}$, 7$-$84 minutes, 2.94$-$6.25\,$R_{\sun}$, and 38.8$-$225.4 Mm.
In COR2 images, the apparent speeds, lifetime, maximal height, and width of the blobs are
336$-$712 km\,s$^{-1}$, 20$-$100 minutes, 4.09$-$7.61\,$R_{\sun}$, and 102.9$-$318.3 Mm.
A positive correlation is found between the lifetime and maximal height, suggesting that blobs with longer lifetime propagate further from the solar surface.
The total duration of blobs is $\ge$8 hr, which is consistent with ongoing flare reconnection 
manifested by persistent ribbon separation in the chromosphere.
The aspect ratio (28$-$37) of the current sheet is probably larger than the critical value,
suggesting that the blobs following the CME are possibly generated by tearing-mode instability of the current sheet.
}

   \authorrunning{Zhang et al.}
   \titlerunning{Multiple blobs following a CME}

   \maketitle

\section{Introduction} \label{intro}
Filaments are cool and dense plasmas floating in the hot and tenuous solar corona \citep[see reviews][and references therein]{mac10}.
The magnetic topology associated with filaments are believed to be sheared arcades \citep{ks57,zes26} and flux ropes \citep{kr74}.
Occasionally, both of them coexist in a single filament \citep{guo10}. The lifetimes of filaments range from days to months.
After losing an equilibrium or being disturbed severely, filaments may become unstable and erupt \citep[e.g.,][]{ant99,chen00,lin00,mo01,tor04,kli06}.
The eruptions of filaments are generally classified into three types: failed eruptions \citep{ji03}, 
partial eruptions \citep{gil00,zqm15,dai22}, and full eruptions \citep{wyp17}.
Failed eruptions are not associated with coronal mass ejections (CMEs), while partial and full eruptions usually generate CMEs.

As shown in the schematic cartoon \citep{lin04}, when a flux rope lifts off and drives a CME, 
a thin current sheet forms between the flux rope and hot post-flare loops \citep{webb03,cia08,sea17}.
As magnetic reconnection proceeds, the current sheet extends and flare ribbons in the chromosphere separate continuously \citep{qiu02,tian16}.
Sometimes, bright and compact blobs, which correspond to radiatively enhanced plasma clumps, 
are detected in the current sheets trailing CMEs \citep{ko03,asai04,lin05,nis10,shq12a,guo13,webb16,chae17,lee20,bem22}.
The blobs show up in series in most cases \citep{shq12a}.
The velocities of these blobs moving outward range from a few hundred to $\ge$1000 km\,s$^{-1}$.
The corresponding reconnection rate in terms of Alfv\'{e}n Mach number ($M_A$) is estimated to be 0.01$-$0.23 \citep{lin05}.
Some of the blobs propagate at constant speeds, while others may accelerate or decelerate \citep{shq12a}.
The lifetimes and apparent widths of the blobs are tens of minutes and 0.14$-$0.35\,$R_{\sun}$ \citep{lee20}.

Magnetohydrodynamic (MHD) numerical simulations have revealed that tearing-mode instability takes place in thin current sheets 
when the aspect ratio ($L/\delta=\sqrt{S_{L}}$) exceeds a threshold (2$\pi$). 
Here, $L$ and $\delta$ denote the length and thickness of a current sheet \citep{fur63}. 
$S_{L}=v_{A}L/\eta$ is the Lundquist number, $v_A$ is the upstream Alfv\'{e}n speed, and $\eta$ is the magnetic diffusivity \citep{lou07}.
It should be pointed out that the critical aspect ratio is dependent on many factors, such as the initial condition, plasma $\beta$, 
and other parameters \citep{bisk00,bha09,ni10,ni12,ni13}. Therefore, the values may range from 20 to 200 in reality.
Tearing-mode instability is referred to as plasmoid instability when entering into the nonlinear stage.
Occurrence of tearing-mode or plasmoid instability gives rise to the formation of magnetic islands or plasmoids \citep{fur63,dra06,ril07,shen11,mei12,lin15}.
Magnetic islands are a set of closed magnetic field lines surrounding O-type null points in two-dimensional (2D) space.
Plenty of numerical simulations have demonstrated that magnetic islands actually correspond to tiny flux ropes formed 
in three-dimensional (3D) current sheets \citep{wyp16,yan22,cgc25}.
The clumped plasma inside these magnetic islands or flux ropes are called plasmoids. 
The observed blobs may correspond to these plasmoids in the current sheets.

Although blobs in the wake of CMEs have been reported in case studies and statistical analysis \citep[e.g.,][]{lin05,shq12a,cx18,lee20}, 
observations of blobs from multiple perspectives are lacking \citep{shq12b}.
In this paper, using multipoint observations, we report detections of blobs in a current sheet trailing a CME on 2023 March 20. 
The paper is organized as follows. Data analysis of multipoint observations are presented in Section~\ref{data}. 
The results are presented in Section~\ref{res}. Discussions and a brief summary are arranged in Section~\ref{dis} and Section~\ref{sum}, respectively.

\begin{table}
	\centering
	\caption{Properties of the instruments used in this study.}
	\label{tab1}
	\begin{tabular}{cccc}
	        \hline
		Instrument & Waveband & Cadence & Pixel Size \\
		                  & ({\AA})        & (sec)        & (arcsec)    \\
		\hline
		GOES & 1$-$8 & 1 & $-$ \\
                 GONG & 6562.8 & 60 &  1.07  \\
		 SDO/AIA & 131, 193, 304 & 12 & 0.60 \\
		 SDO/HMI & 6173 & 45 & 0.60 \\
		 SOHO/LASCO-C2 & WL & $\sim$420 & 11.60 \\
		 SOHO/LASCO-C3 & WL & $-$ & 53.50 \\
                  STA/COR2 & WL & $\sim$600 & 56.65 \\
		\hline
	\end{tabular}
\end{table}

\section{Data analysis of multipoint observations} \label{data}
\begin{figure}
\centering
\includegraphics[width=0.7\columnwidth]{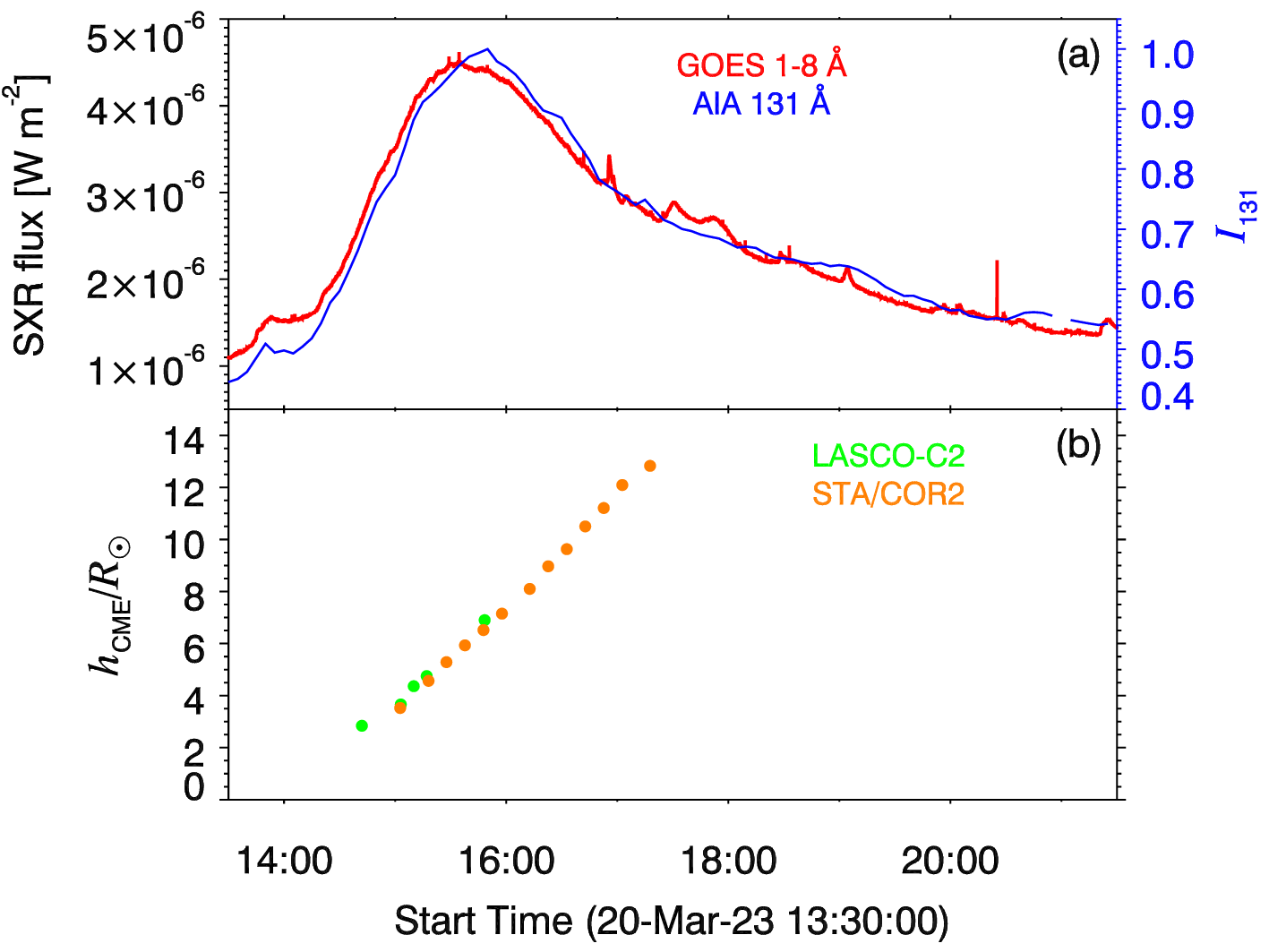}
    \caption{(a) Light curves of the C4.1 class flare in AIA 131 {\AA} (blue line) and GOES 1$-$8 {\AA} (red line).
    (b) Time variations of the CME height observed by LASCO-C2 (green dots) and STA/COR2 (orange dots).}
    \label{fig1}
\end{figure}

On 2023 March 20, a long filament residing in NOAA active region (AR) 13258 (N32E26) in the north hemisphere was observed in H$\alpha$
line center by the Global Oscillation Network Group (GONG) and in 304 {\AA} by the Atmospheric Imaging Assembly \citep[AIA;][]{lem12}
on board the Solar Dynamics Observatory \citep[SDO;][]{pes12}.
AIA takes full-disk images in 7 extreme ultraviolet (EUV) wavelengths (94, 131, 171, 193, 211, 304, 335 {\AA}) and two UV wavelengths (1600 and 1700 {\AA}).
The C4.1 flare generated by the filament eruption was observed in 131 {\AA} by SDO/AIA and 
in 1$-$8 {\AA} by the Geostationary Operational Environmental Satellite (GOES) spacecraft.
In Figure~\ref{fig1}, the top panel shows light curves of the flare in 131 {\AA} (blue line) and 1$-$8 {\AA} (red line), respectively.
The flare starts at $\sim$14:15 UT, peaks at $\sim$15:13 UT, and stops at $\sim$16:30 UT.

The filament eruption also produced a large CME\footnote{https://cdaw.gsfc.nasa.gov/CME\_list/UNIVERSAL\_ver2/2023\_03/univ2023\_03.html}
observed by the C2 and C3 coronagraphs of the Large Angle Spectroscopic Coronagraph \citep[LASCO;][]{bru95} 
on board the Solar and Heliospheric Observatory (SOHO) 
and the COR2 coronagraph of the Sun Earth Connection Coronal and Heliospheric Investigation \citep[SECCHI;][]{how08}
on board the ahead Solar TErrestrial RElations Observatory \citep[STEREO;][]{kai08}.
The separation angle between ahead STEREO (hereafter STA) and the Sun-Earth connection was $\sim$12$\degr$.
There was a data gap during 18:47$-$19:22 UT in LASCO-C2 images.
Full-disk line-of-sight (LOS) magnetograms of the photosphere were observed by the Helioseismic and Magnetic Imager \citep[HMI;][]{sch12} on board SDO.
The 3D magnetic configuration supporting the filament is reconstructed using nonlinear force-free field (NLFFF) extrapolation based on vector magnetograms 
of the photosphere observed by HMI before eruption.
We use the magnetofrictional relaxation method \citep{guo16}, which is implemented in the MPI-AMRVAC framework \citep{kep23}.
The grid points of the domain of extrapolation are 360$\times$140$\times$140 with a grid size of $\sim$4$\arcsec$.
Hence, the true size of the domain containing AR 13258 is 1044$\times$406$\times$406 Mm$^3$.
The properties of instruments used in this study are summarized in Table~\ref{tab1}.

\section{Results} \label{res}
\subsection{Partial filament eruption and C4.1 class flare} \label{prim} 
\begin{figure}
\centering
	\includegraphics[width=0.9\columnwidth]{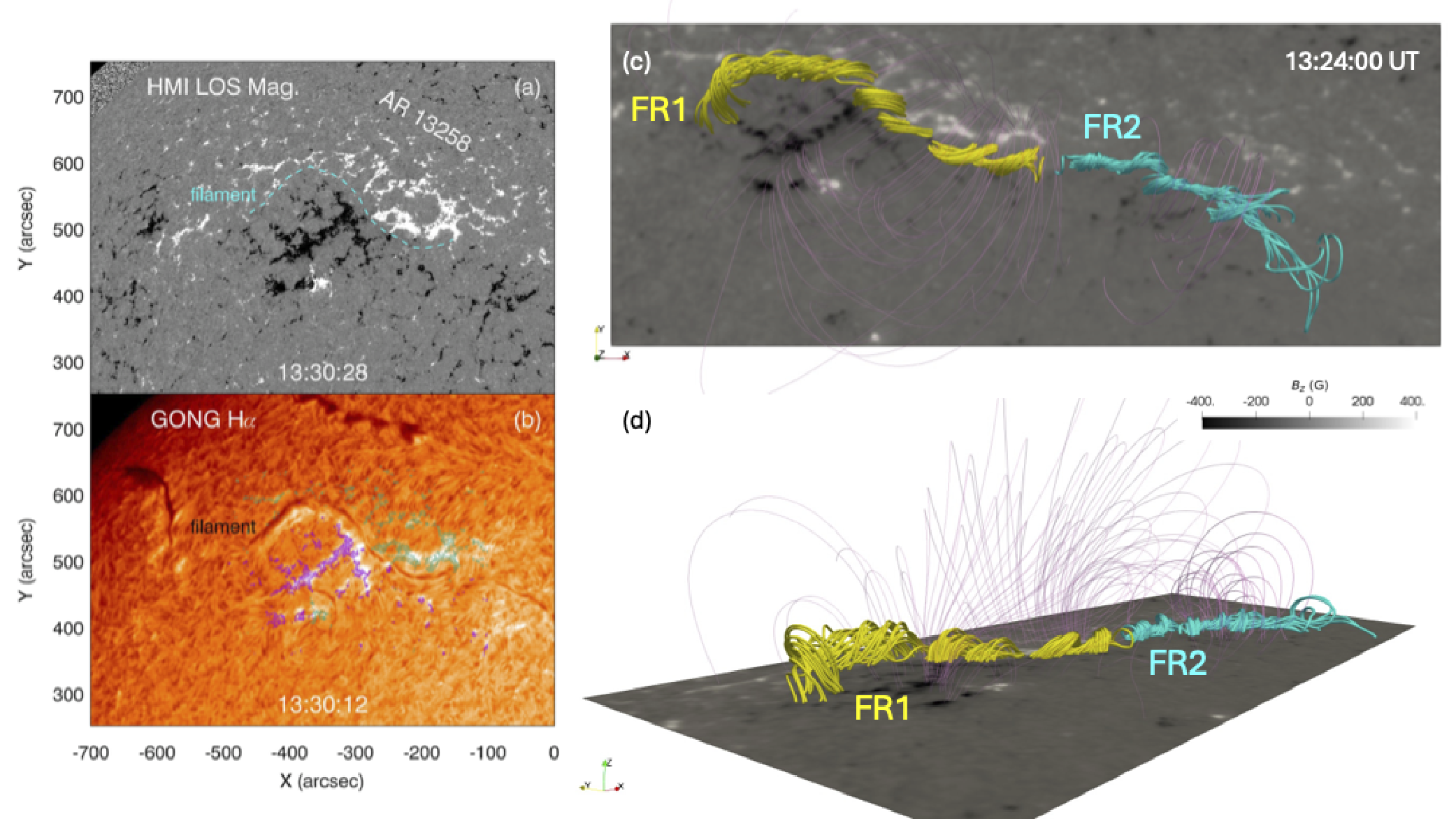}
    \caption{Left panels: HMI LOS magnotogram and GONG H$\alpha$ image at 13:30 UT.
    Right panels: top view and side view of 3D magnetic configuration supporting the filament at 13:24 UT.
    Yellow and turquoise lines represent magnetic field lines of FR1 and FR2, respectively.
    Mauve lines represent overlying field lines above the filament.}
    \label{fig2}
\end{figure}

In Figure~\ref{fig2}, the left panels show HMI LOS magnetogram and GONG H$\alpha$ image at 13:30 UT.
In panel (a), the cyan dashed line represents the spine of sigmoidal filament with a length of $\sim$381$\arcsec$ along the polarity inversion line (PIL) of AR 13258.
In panel (b), the contours of positive and negative magnetic field strength are overlaid with green and purple lines, respectively.
The eastern part of the filament is darker and wider than the western part.
The right panels show top view and side view of 3D magnetic configuration supporting the filament at 13:24 UT.
It is found that the filament is composed of two flux ropes along the PIL: FR1 with yellow lines and FR2 with turquoise lines.
The mauve lines represent overlying field lines above the filament.

\begin{figure}
\centering
	\includegraphics[width=0.9\columnwidth]{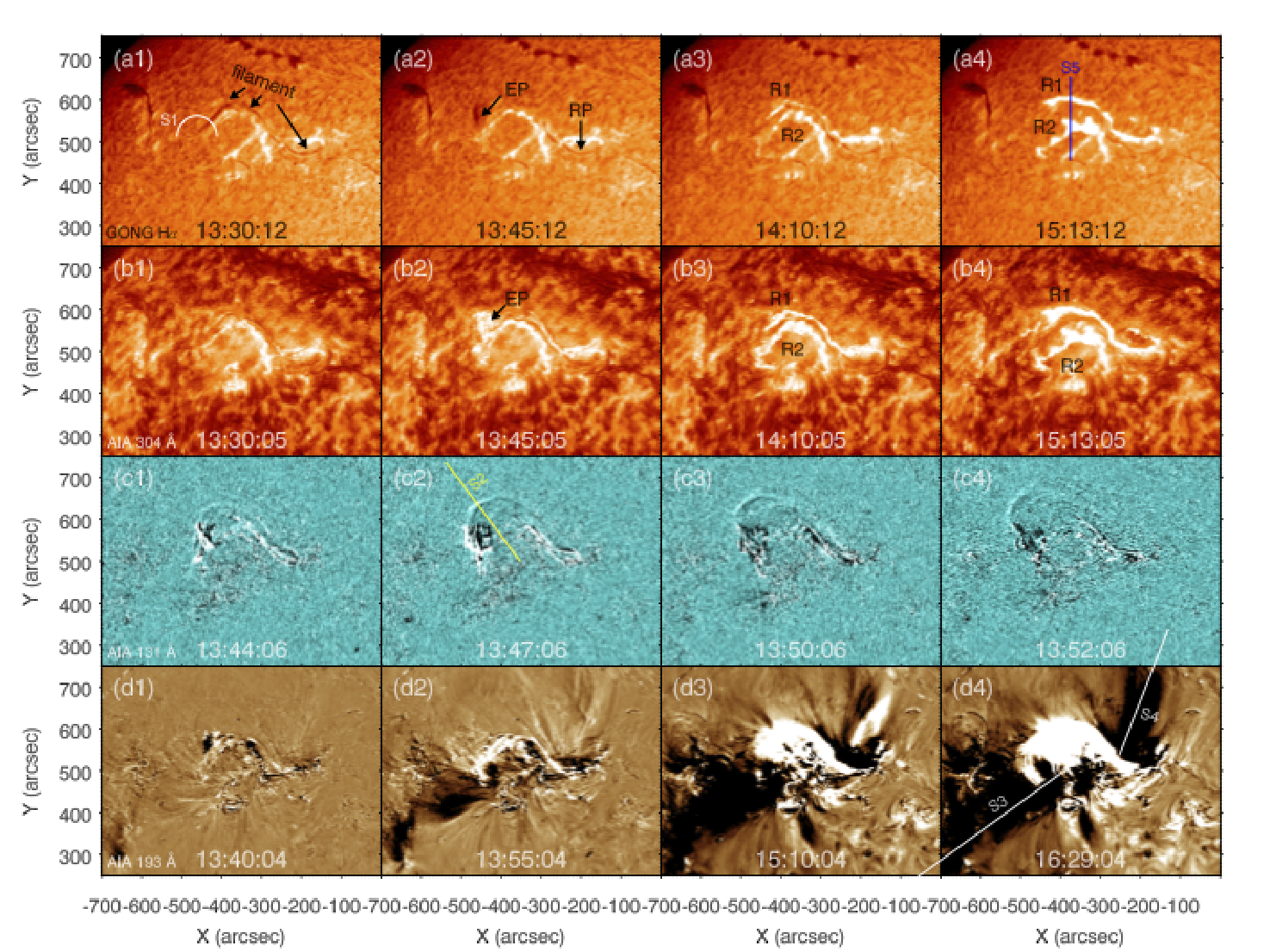}
    \caption{(a1)-(b4) Filament eruption and flare ribbons in GONG H$\alpha$ and AIA 304 {\AA} images.
              The black arrows point to the eruptive part (EP) and remaining part (RP) of the sigmoidal filament.
              R1 and R2 denote the north and south flare ribbons, respectively.
              (c1)-(c4) Eruption of the bright arc in AIA 131 {\AA} running-difference images.
              (d1)-(d4) Twin dimming at two sides of the flare region.
              Five slices (S1-S5) are used to investigate the filament eruption, twin dimming, and ribbon separation.
              An online animation of the GONG H$\alpha$ and AIA 304 {\AA} images is available. 
              The $\sim$6 s animation covers from 13:30 UT to 15:30 UT.}
    \label{fig3}
\end{figure}

\begin{figure}
\centering
	\includegraphics[width=0.6\columnwidth]{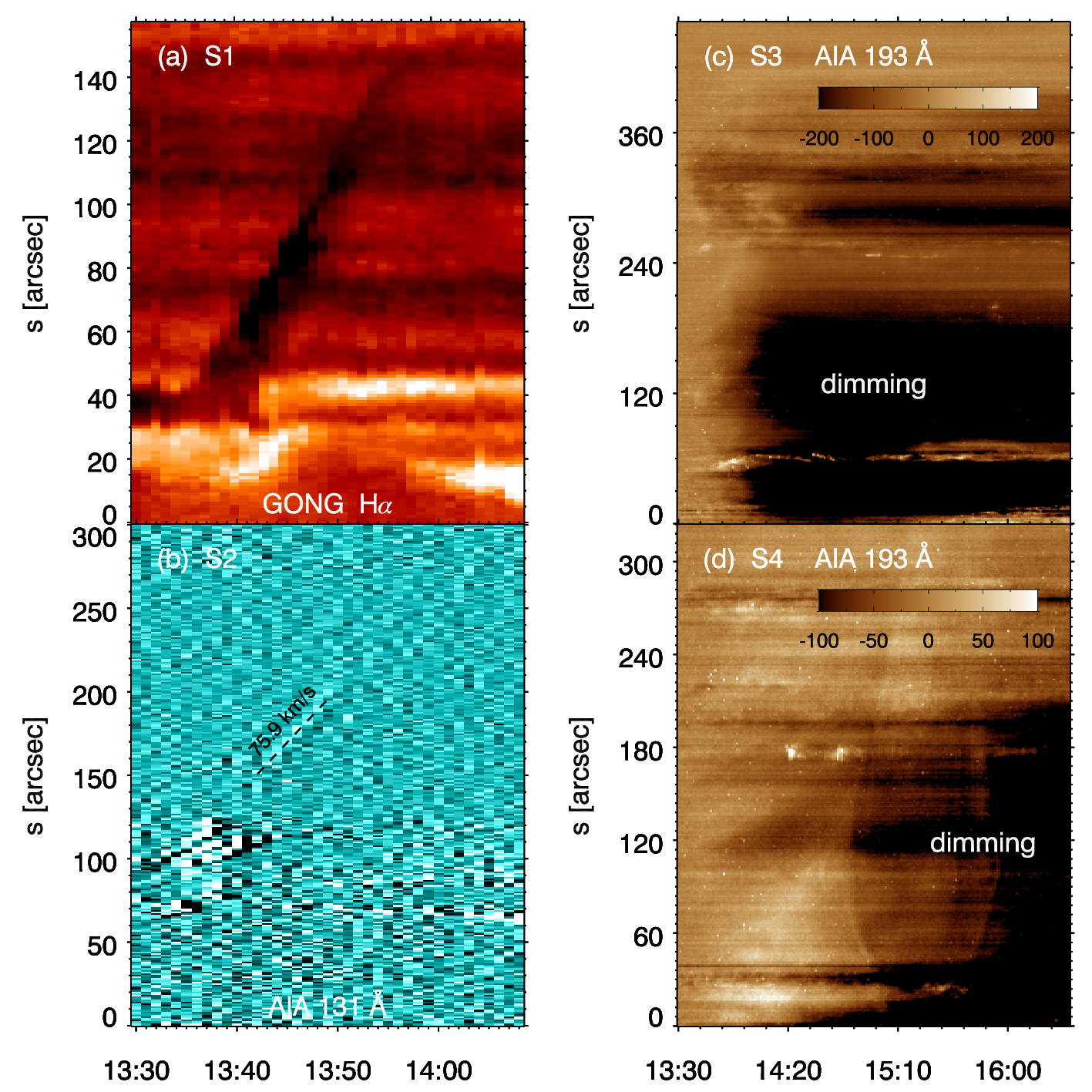}
    \caption{Time-distance diagrams of S1 in H$\alpha$, S2 in AIA 131 {\AA}, S3 in AIA 193 {\AA}, and S4 in AIA 193 {\AA}.
		In panel (b), the rising speed ($\sim$75.9 km\,s$^{-1}$) of bright arc is labeled.
		In panels (c)-(d), the dark dimmings are labeled.}
    \label{fig4}
\end{figure}

In Figure~\ref{fig3}, the top two rows show evolutions of the filament and flare ribbons in H$\alpha$ and 304 {\AA} images (see also the online animation).
It is seen that the filament undergoes a partial eruption, i.e., the eruptive part (hereafter EP) rises up and the remaining part (hereafter RP) stays behind.
EP and RP are roughly associated with FR1 and FR2 in Figure~\ref{fig2}.
The eruption of EP generates a pair of flare ribbons, the north ribbon R1 and south ribbon R2 (see panels (a3) and (b3)).
In panel (a1), a curved slice S1 is selected to go across the filament. 
Time-distance diagram of S1 in H$\alpha$ is displayed in Figure~\ref{fig4}(a).
It is obvious that the filament starts to lift off at $\sim$13:35 UT and moves eastward till $\sim$13:55 UT when the EP fades out.

The partial eruption is also detected in AIA 131 {\AA}. Figure~\ref{fig3}(c1)-(c4) show running-difference images in 131 {\AA} during 13:44$-$13:52 UT.
A bright arc moves in the northeast direction as the filament erupts. In panel (c2), a yellow slice S2 with a length of 300$\arcsec$ is selected to explore the eruption.
Time-distance diagram of S2 is displayed in Figure~\ref{fig4}(b). The black dashed line indicates eruption of the bright arc at a speed of $\sim$75.9 km\,s$^{-1}$
during 13:41$-$13:50 UT. The bright arc seen in 131 {\AA} may stand for overlying field lines above the cool filament. 

As the filament moves forward, a pair of twin dimming come out at two sides of the flare \citep{ste97,ver25}.
Figure~\ref{fig3}(d1)-(d4) show base-difference images in AIA 193 {\AA} during 13:40$-$16:29 UT.
The dimming first appears to the east of flare region (panel (d2)). 
As time goes on, the eastern dimming spreads out quickly and the western dimming becomes manifest (panel (d3)).
In order to investigate the evolutions of twin dimming, we select two slices (S3 and S4) with total lengths of 462$\arcsec$ and 324$\arcsec$ in panel (d4).
Time-distance plots of S3 and S4 are displayed in the right panels of Figure~\ref{fig4}. It's obvious that the eastern dimming is earlier and darker than the western dimming.

\begin{figure}
\centering
	\includegraphics[width=0.8\columnwidth]{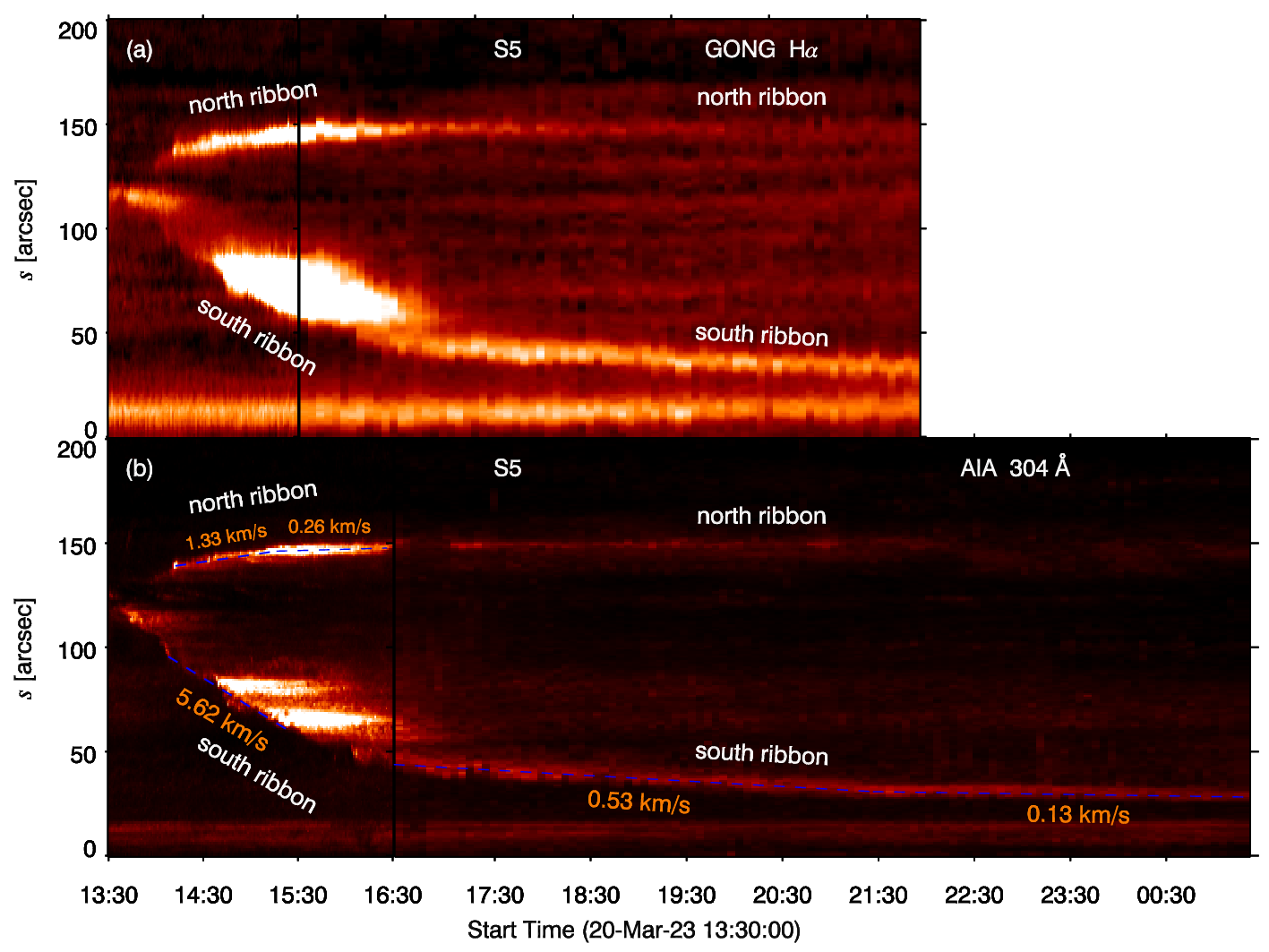}
    \caption{Time-distance diagrams of S5 in GONG H$\alpha$ (a) and AIA 304 {\AA} (b).
                  Separation speeds of the north and south ribbons are labeled.}
    \label{fig5}
\end{figure}

As seen in top two rows of Figure~\ref{fig3}, the conjugate flare ribbons (R1 and R2) separate from each other as the flare proceeds.
To investigate the separation, we select a vertical slice S5 with a length of 200$\arcsec$ in Figure~\ref{fig3}(a4).
Time-distance diagrams of S5 in H$\alpha$ and 304 {\AA} are plotted in the top and bottom panels of Figure~\ref{fig5}.
It is clear that the ribbons separate from each other quickly from the beginning of flare at 14:15 UT until $\sim$16:30 UT.
The speeds of separation are 5.62 and 0.26$-$1.33 km\,s$^{-1}$ for the south and north ribbons, respectively.
After 16:30 UT, the brightness of flare ribbons reduces sharply.
The north ribbon comes to a halt, while the south ribbon continues to move at speeds of 0.13$-$0.53 km\,s$^{-1}$,
implying that the magnetic reconnection rate decreases with time.
\citet{lyn04} performed a numerical simulation of a flux rope eruption in the breakout model.
As the flux rope rises, both the height of post-flare loops and ribbon separation increase with time.
The speed of separation decreases quickly from $\sim$43 km\,s$^{-1}$ in the early phase to $\sim$7 km\,s$^{-1}$ in the later phase.
The whole eruption lasts for $\sim$9 hr.
Although the ribbon separation speeds in their simulation are much larger than the observed values, they have the same trend.

\begin{figure}
\centering
	\includegraphics[width=0.8\columnwidth]{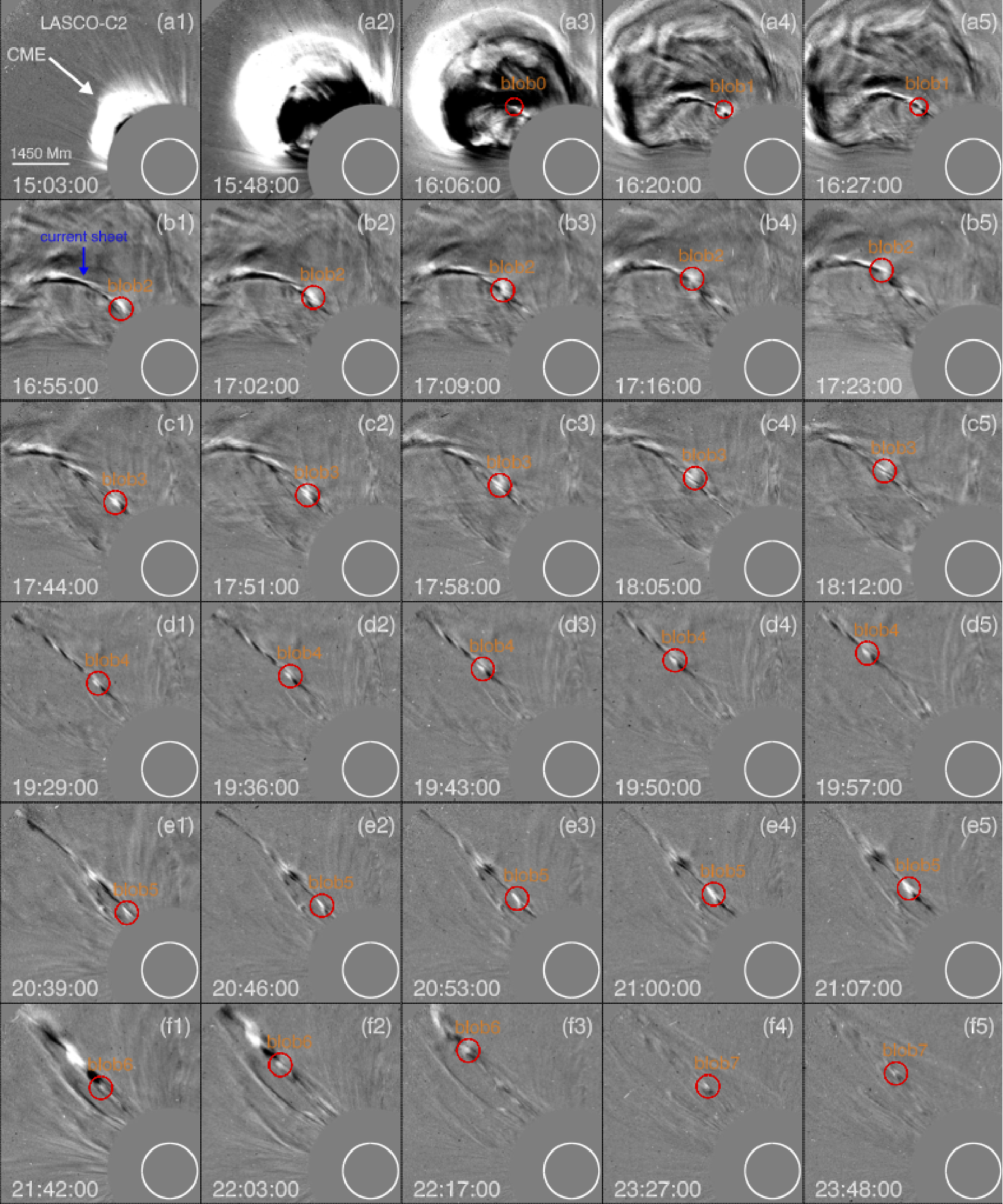}
    \caption{CME, current sheet, and blobs observed by LASCO-C2 during 15:03$-$23:48 UT.
		The white arrow points to the bright CME front in panel (a1). 
		The blue arrow points to the current sheet in panel (b1). 
		The blobs are encircled by red circles.
                 An online animation of the CME and blobs in LASCO-C2 running-difference images is available. 
                 The $\sim$3 s animation covers from 14:21 to 23:48 UT.}
    \label{fig6}
\end{figure}

\subsection{Current sheet and blobs after the CME} \label{blob}
The partial filament eruption generates a wide CME propagating in the northeast direction.
Figure~\ref{fig6} shows running-difference images of LASCO-C2 during 15:03$-$23:48 UT with a field of view of 7000$\arcsec\times7000\arcsec$.
The white circles indicate the solar limb at 1\,$R_{\sun}$. In panel (a1), the white arrow points to the bright CME front.
Time evolution of the heliocentric distance of the CME front ($h_{\mathrm{CME}}$) is plotted with green dots in Figure~\ref{fig1}(b).
The value of $h_{\mathrm{CME}}$ increases from $\sim$2.84\,$R_{\sun}$ at 14:42 UT to $\sim$6.90\,$R_{\sun}$ at 15:48 UT.
A linear fitting results in an apparent speed of $\sim$727.5 km\,s$^{-1}$. 
Assuming that the CME propagates radially, the true speed may exceed 1600 km\,s$^{-1}$ according to the source location in AR 13258.
From Figure~\ref{fig6} and the online animation, it is noteworthy that a thin current sheet appears and extends up in the wake of CME.
The post-CME current sheet is a little curved instead of straight, like the one on 2021 February 13 \citep{bem22}.
The direction of current sheet changes slightly as the CME moves forward, but mostly in the northeast direction.
Meanwhile, a series of bright and compact blobs flow out of the current sheet.  
In LASCO-C2 running-difference images, a total of eight blobs (blob0-blob7) are identified and encircled with red circles during 16:00$-$23:59 UT.
The first one (blob0) appears at 16:06 UT, which is $\sim$53 minutes later than the flare peak time.

\begin{figure}
\centering
	\includegraphics[width=0.85\columnwidth]{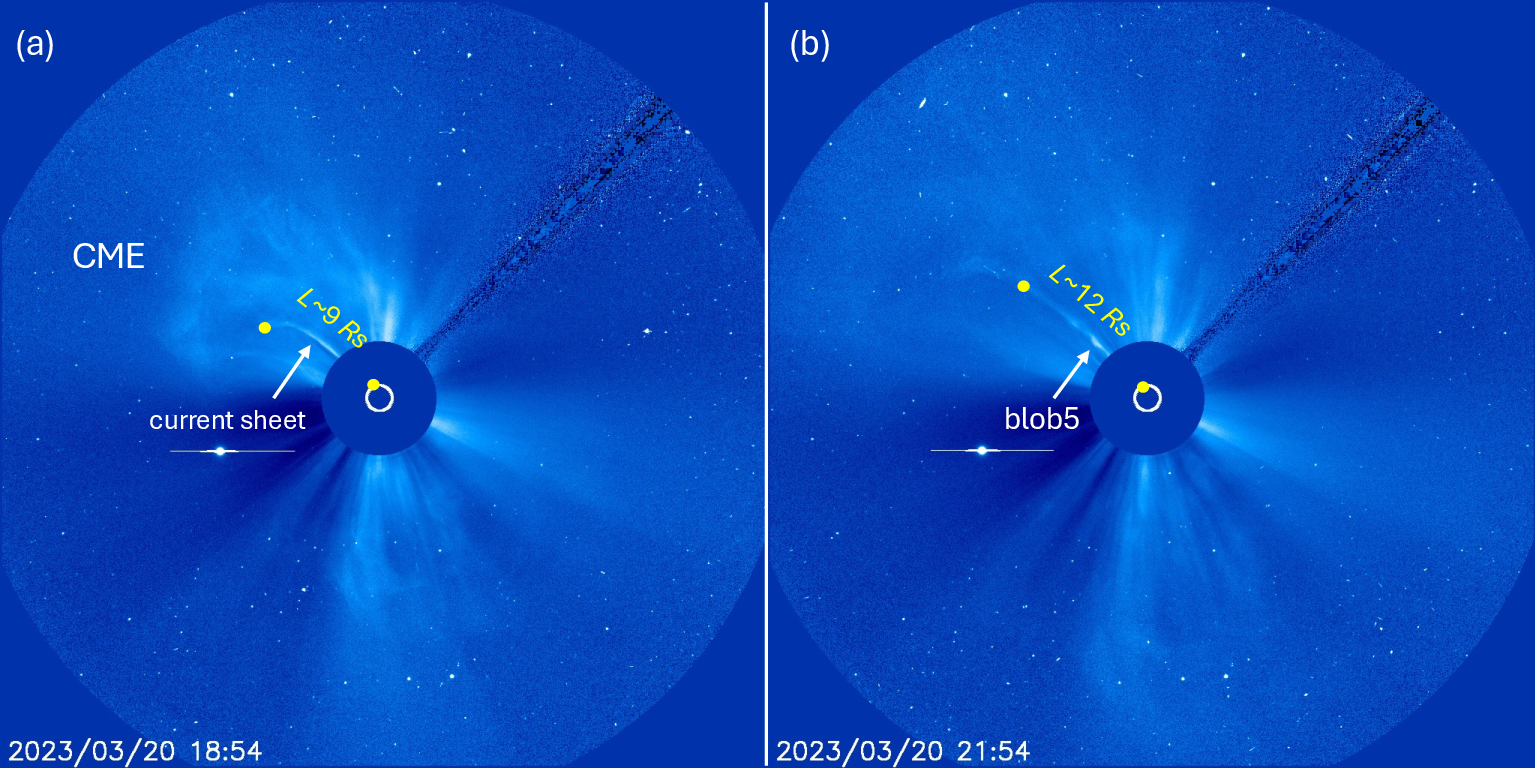}
    \caption{CME and blob5 within the current sheet observed by LASCO-C3 (original images) at 18:54 UT (a) and 21:54 UT (b).
		The white arrows point to the current sheet and blob5, respectively.
		The yellow dots represent the endpoints of current sheet, whose lengths ($\sim$9\,$R_{\sun}$ and $\sim$12\,$R_{\sun}$) are labeled.}
    \label{fig7}
\end{figure}

The CME and current sheet are also observed by LASCO-C3, although there are only two images during the whole event.
Figure~\ref{fig7} shows original images of LASCO-C3 at 18:54 UT (a) and 21:54 UT (b).
In the left panel, the white arrow points to the sharp, curved current sheet behind the CME, which looks the same as that observed by LASCO-C2.
Two yellow dots represent the endpoints of current sheet with an apparent length of $\sim$9\,$R_{\sun}$.
In the right panel, the current sheet becomes longer with an apparent length of $\sim$12\,$R_{\sun}$.
The white arrow points to blob5, which has a heliocentric distance of $\sim$6.0\,$R_{\sun}$.

\begin{figure}
\centering
	\includegraphics[width=0.7\columnwidth]{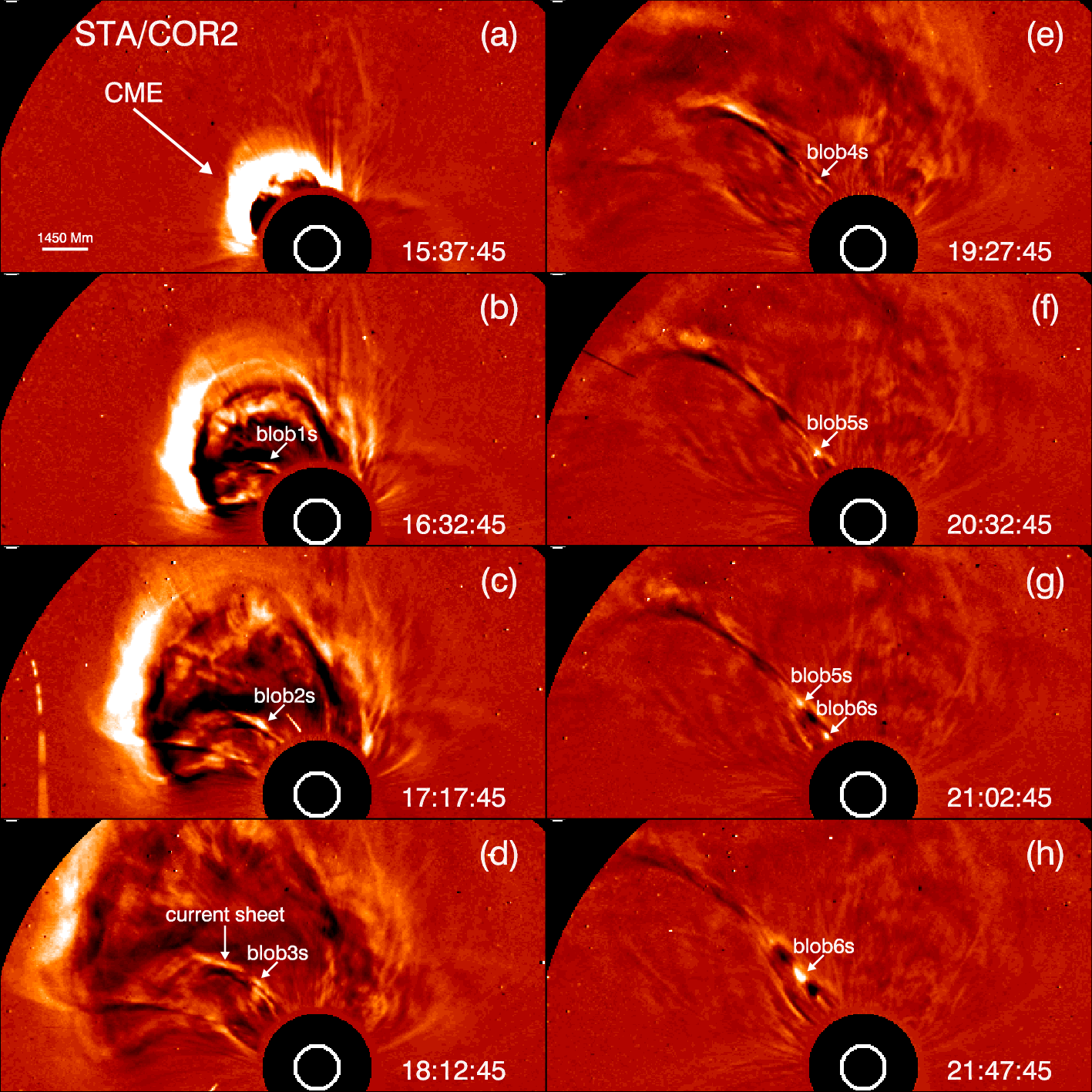}
    \caption{CME, current sheet, and blobs observed by STA/COR2.
                The long arrow points to the bright CME front in panel (a).
                The vertical arrow points to the current sheet in panel (d).
                The short arrows point to the blobs when they first appear.
                An online animation of the CME and blobs in COR2 running-difference images is available. 
                The $\sim$2 s animation covers from 15:02 UT to 23:48 UT.}
    \label{fig8}
\end{figure}

The CME, current sheet, and blobs are also detected by STA/COR2 from another viewpoint. 
Figure~\ref{fig8} shows running-difference images of STA/COR2 during 15:37$-$21:47 UT (see also the online movie).
In panel (a), the long arrow points to the CME front. Time variation of the CME front height is plotted with orange dots in Figure~\ref{fig1}(b).
The value of $h_{\mathrm{CME}}$ increases from $\sim$3.52\,$R_{\sun}$ at 15:02 UT to $\sim$12.83\,$R_{\sun}$ at 17:17 UT.
A linear fitting gives rise to an apparent speed of $\sim$812 km\,s$^{-1}$.
In panels (b)-(h), the short arrows point to the six blobs (blob1s-blob6s) identified in COR2 images (the last blob blob7s is not shown).
Since the separation angle ($\sim$12$\degr$) between STA and Sun-Earth connection is small, 
the shapes of current sheet and blobs seen in LASCO and COR2 are comparable.
It is interesting that multiple blobs coexist along the current sheet sometimes (see panel (g)).

\begin{figure}
\centering
	\includegraphics[width=0.7\columnwidth]{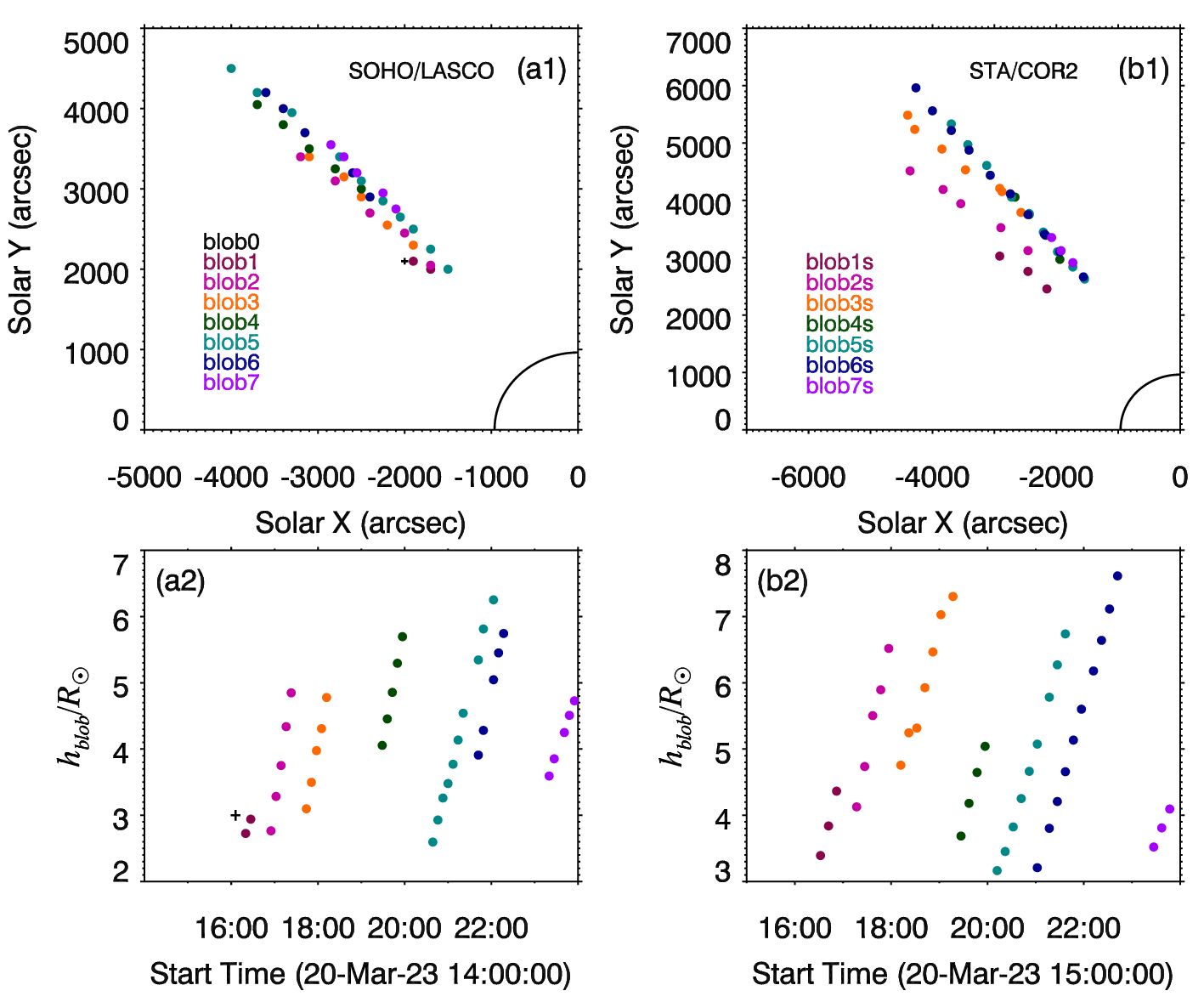}
    \caption{Top panels: Trajectories of the blobs observed by LASCO-C2 and STA/COR2 coronagraphs.
		Bottom panels: Time evolutions of the heliocentric distances of the corresponding blobs.}
    \label{fig9}
\end{figure}

\begin{figure}
\centering
	\includegraphics[width=0.7\columnwidth]{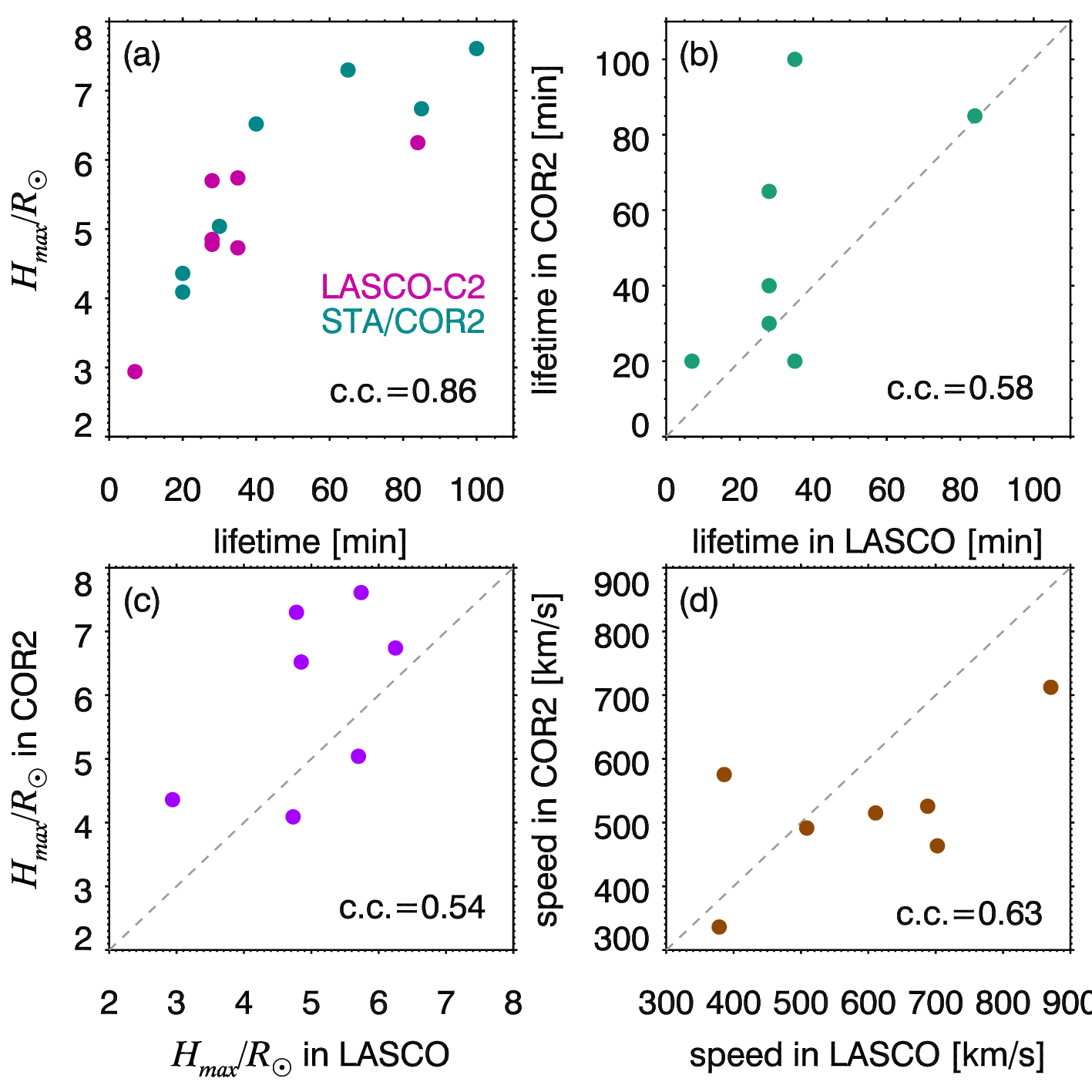}
    \caption{(a) Scatter plot between the lifetimes and maximal heights of the blobs.
    (b)-(d) Scatter plots of lifetimes, maximal heights, and speeds of the blobs observed by LASCO-C2 and STA/COR2.
    Correlation coefficients are labeled in each panel.}
    \label{fig10}
\end{figure}

In Figure~\ref{fig9}, trajectories of these blobs observed in LASCO and COR2 are drawn with colored dots in the top panels.
Time evolutions of the heliocentric distances of the corresponding blobs are drawn with colored dots in the bottom panels.
Note that blob0 is identified only at one moment and could not be tracked continuously. Therefore, the position and trajectory of blob0 are drawn with black pluses.
It is obvious that the height evolutions of blob1$-$blob7 are roughly concurrent with those of blob1s$-$blob7s.
In other words, blob1$-$blob7 roughly have a one-to-one correspondence with blob1s$-$blob7s.
Due to the difference in time cadence of the two coronagraphs, the times of appearance and disappearance of blobs are not exactly the same in the coronagraphs.
The maximal heights of these blobs are between $\sim$2.94\,$R_{\sun}$ and $\sim$6.25\,$R_{\sun}$ with a mean value of $\sim$5.0\,$R_{\sun}$ in LASCO-C2 images,
while their heights reach up to 4.09$-$7.61\,$R_{\sun}$ with a mean value of $\sim$5.95\,$R_{\sun}$ in COR2 images.
The difference in height may result from different FOVs of two coronagraphs.
Figure~\ref{fig10}(b)-(d) show scatter plots of lifetimes, maximal heights, and speeds of the blobs observed by LASCO-C2 and STA/COR2, respectively.
The correlation coefficients are between 0.54 and 0.63, indicating that parameters of the blobs detected in two coronagraphs have positive correlations.

It is revealed from Figure~\ref{fig9}(a2)-(b2) that most of the blobs propagate with nearly constant speeds. 
No acceleration or deceleration are found for these blobs. The apparent speeds and lifetime of these blobs are calculated and listed in Table~\ref{tab2}.
It is seen that the speeds range from $\sim$378 to $\sim$871 km\,s$^{-1}$ with a mean value of $\sim$592 km\,s$^{-1}$ in the FOV of LASCO,
while the speeds are between $\sim$336 and $\sim$712 km\,s$^{-1}$ with a mean value of $\sim$517 km\,s$^{-1}$ in the FOV of COR2.
The lifetime fall in a range of 7$-$84 minutes with a mean value of $\sim$35.0 minutes in LASCO images,
while the lifetime is between 20 and 100 minutes with a mean value of $\sim$51.4 minutes in COR2 images.
Among them, blob5 has the largest lifetime ($\sim$84 minutes) and maximal height ($\sim$6.25\,$R_{\sun}$) in LASCO images, 
while blob6s has the largest lifetime ($\sim$100 minutes) and maximal height ($\sim$7.61\,$R_{\sun}$) in COR2 images.
The total duration of blob formation and propagation are $\ge$8 hrs, 
which is in agreement with the fact that flare reconnection in the current sheet persists for $\sim$12 hrs manifested by continuous ribbon separation (Figure~\ref{fig5}(b)).

It is emphasized that the measured heights and speeds of blobs are apparent values as a result of projection effect.
Considering the source region of the filament eruption at AR 13258, the true heights and speeds of blobs should be much larger after deprojection.
Besides, all the blobs move outward rather than sunward. 
Since the corona below 2.5\,$R_{\sun}$ is shaded by the occulting disks of WL coronagraphs, the blobs moving sunward could not be detected.
\citet{ril07} carried out numerical simulations to investigate the current sheet and blobs in the wake of a CME.
Both outwardly-propagating and inwardly-propagating blobs are created. 
The height of main X-point is just around 2.5\,$R_{\sun}$ (see their Fig. 3).
\citet{gui16} performed an ultrahigh-resolution 2.5D numerical simulation of a flux rope eruption in the breakout model \citep{ant99}.
Likewise, both outward moving and sunward-moving magnetic islands are created within the flare current sheet beneath the CME core.
The radial position of main X-null separating the outward and sunward islands varies between 1.30 and 1.45\,$R_{\sun}$.
Similarly, the heights of main X-point in current sheets below CME flux ropes could be much lower than 2.5\,$R_{\sun}$ \citep{mei12,mei17}.
Therefore, the sunward moving islands in simulations could not be observed by LASCO and COR2 coronagraphs.
Since ground-based coronagraphs are capable of observing inner corona, such as the K-coronagraph \citep[K-Cor;][]{tom16} (1.05$-$3.0\,$R_{\sun}$)
and Spectral Imaging Coronagraph \citep[SICG;][]{hf25} (1.05$-$2.0\,$R_{\sun}$), sunward-moving blobs could hopefully be detected by ground-based coronagraphs.

Figure~\ref{fig10}(a) shows a scatter plot between the lifetimes and maximal heights of the blobs.
The magenta and lake blue dots represent blobs observed by LASCO-C2 and STA/COR2, respectively.
A positive correlation between the two parameters is found with a correlation coefficient of $\sim$0.86,
indicating that blobs with longer lifetimes go further from the solar surface.

\begin{figure}
\centering
	\includegraphics[width=0.7\columnwidth]{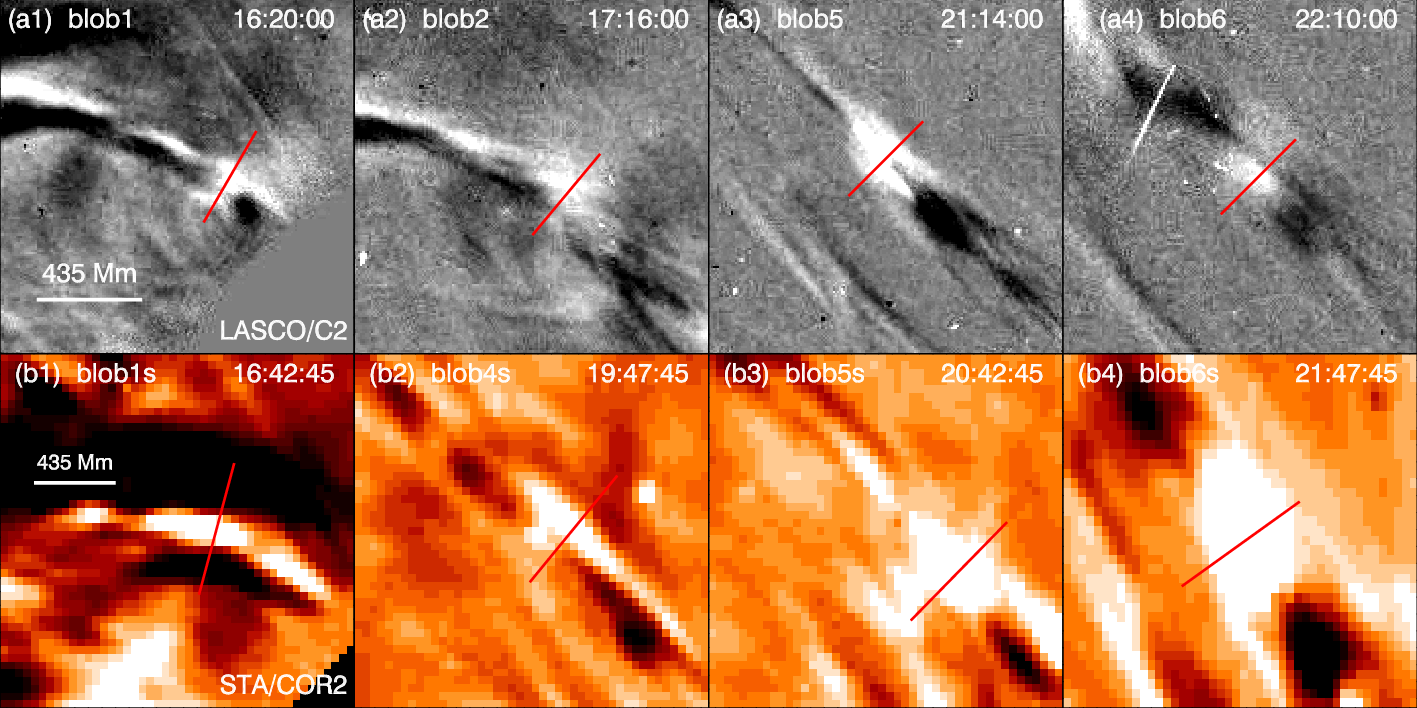}
    \caption{Close-up images of four blobs in LASCO-C2 images (top panels) and four blobs in COR2 images (bottom panels).
              The red slices across these blobs are used to calculate their widths.}
    \label{fig11}
\end{figure}

\begin{figure}
\centering
	\includegraphics[width=0.7\columnwidth]{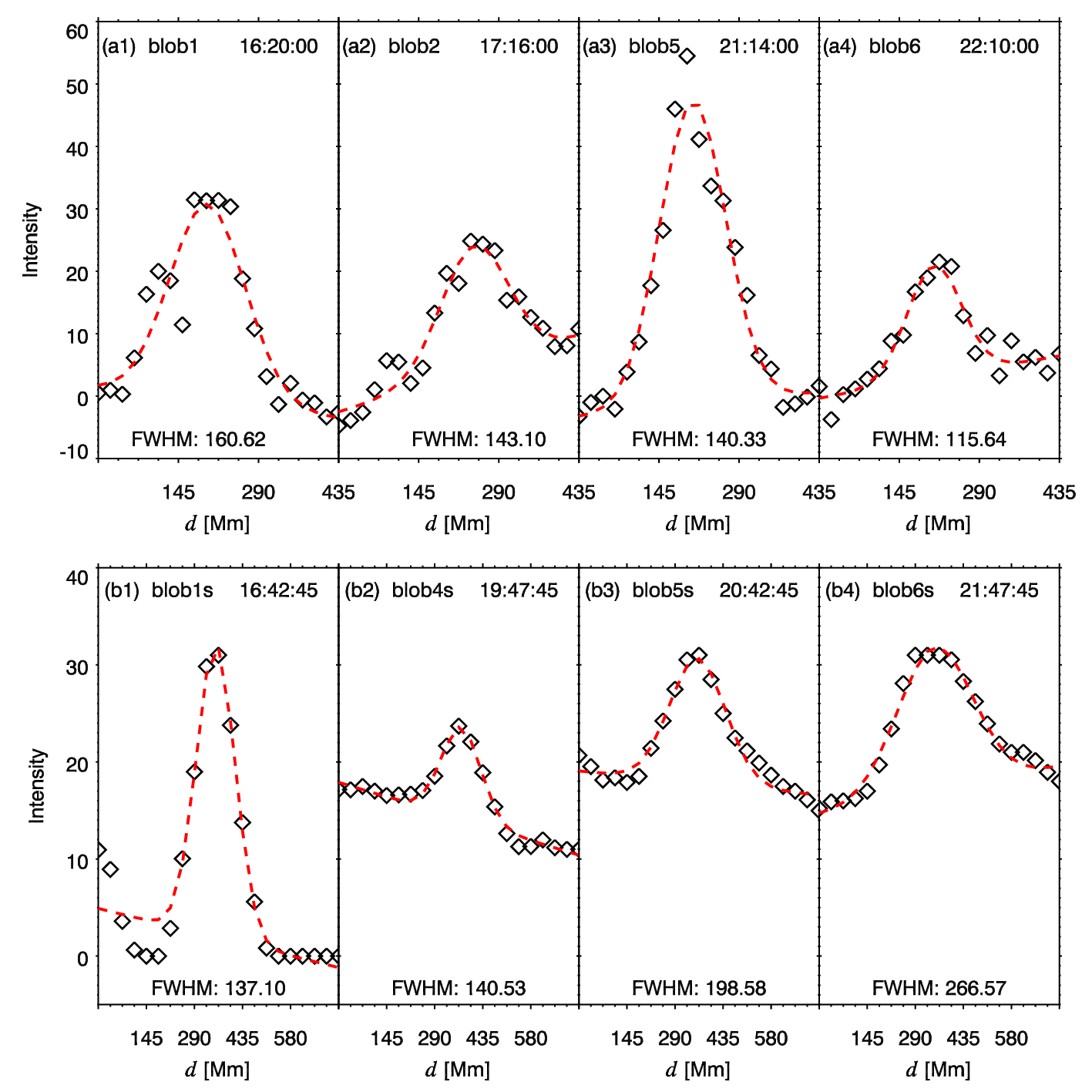}
    \caption{Intensity distributions (black diamonds) along the red slices across the blobs in Figure~\ref{fig11}.
		Red dashed lines stand for results of Gaussian fitting. The values of FWHM in unit of Mm are labeled.}
    \label{fig12}
\end{figure}

In order to calculate the widths of blobs, we derive the intensity distribution across each blob and perform a Gaussian fitting.
Full width at half maximum (FWHM) of the fitted profile is considered as the blob width.
In Figure~\ref{fig11}, the top panels show close-up images of four blobs (blob1, blob2, blob5, and blob6) observed by LASCO-C2.
The bottom panels show close-up images of four blobs (blob1s, blob4s, blob5s, and blob6s) observed by STA/COR2.
The red lines are slices across the blobs to derive their intensity distributions, which are drawn with black diamonds in Figure~\ref{fig12}.
The results of Gaussian fitting are superposed with red dashed lines and the associated values of FWHM are displayed at the bottom of each panel. 

\begin{table}
	\centering
	\caption{Physical properties of blobs in the current sheet, 
		including the width, lifetime, speed, and maximal heliocentric height ($H_{max}$).}
	\label{tab2}
	\begin{tabular}{cccccc}
	         \hline
		Instrument & Blob & Width & Lifetime & Speed            & $H_{max}$ \\
		           &    & (Mm)  & (min)      & (km\,s$^{-1}$) & ($R_{\sun}$) \\
		\hline
                  LASCO-C2 & blob0 & 119.55 & $-$ &  $-$ & 3.00  \\
		 LASCO-C2 & blob1 & 61.87$-$160.62 &   7 &  386.0 & 2.94  \\
		 LASCO-C2 & blob2 & 77.71$-$188.14 & 28 &  870.9 & 4.85  \\
		 LASCO-C2 & blob3 & 78.74$-$225.40 & 28 &  702.5 & 4.78  \\
		 LASCO-C2 & blob4 & 63.69$-$102.32 & 28 &  688.1 & 5.70  \\
		 LASCO-C2 & blob5 & 38.76$-$146.01 & 84 &  508.6 & 6.25  \\
		 LASCO-C2 & blob6 & 119.10$-$166.81 & 35 &  610.8 & 5.74  \\
		 LASCO-C2 & blob7 & 83.44$-$106.20 & 35 &  378.6 & 4.73  \\
		 \hline
		 STA/COR2 & blob1s & 137.10$-$147.53 & 20 &  575.4 & 4.36  \\
		 STA/COR2 & blob2s & 223.79$-$292.13 & 40 &  712.4 & 6.52  \\
		 STA/COR2 & blob3s & 141.29$-$318.34 & 65 &  463.5 & 7.30  \\
		 STA/COR2 & blob4s & 133.08$-$182.32 & 30 &  525.6 & 5.04  \\
		 STA/COR2 & blob5s & 152.17$-$268.71 & 85 &  491.4 & 6.74  \\
		 STA/COR2 & blob6s & 189.72$-$310.82 & 100 &  515.1 & 7.61  \\
		 STA/COR2 & blob7s & 102.94$-$137.61 & 20 &  336.1 & 4.09  \\
		\hline
	\end{tabular}
\end{table}

\begin{figure}
\centering
	\includegraphics[width=0.5\columnwidth]{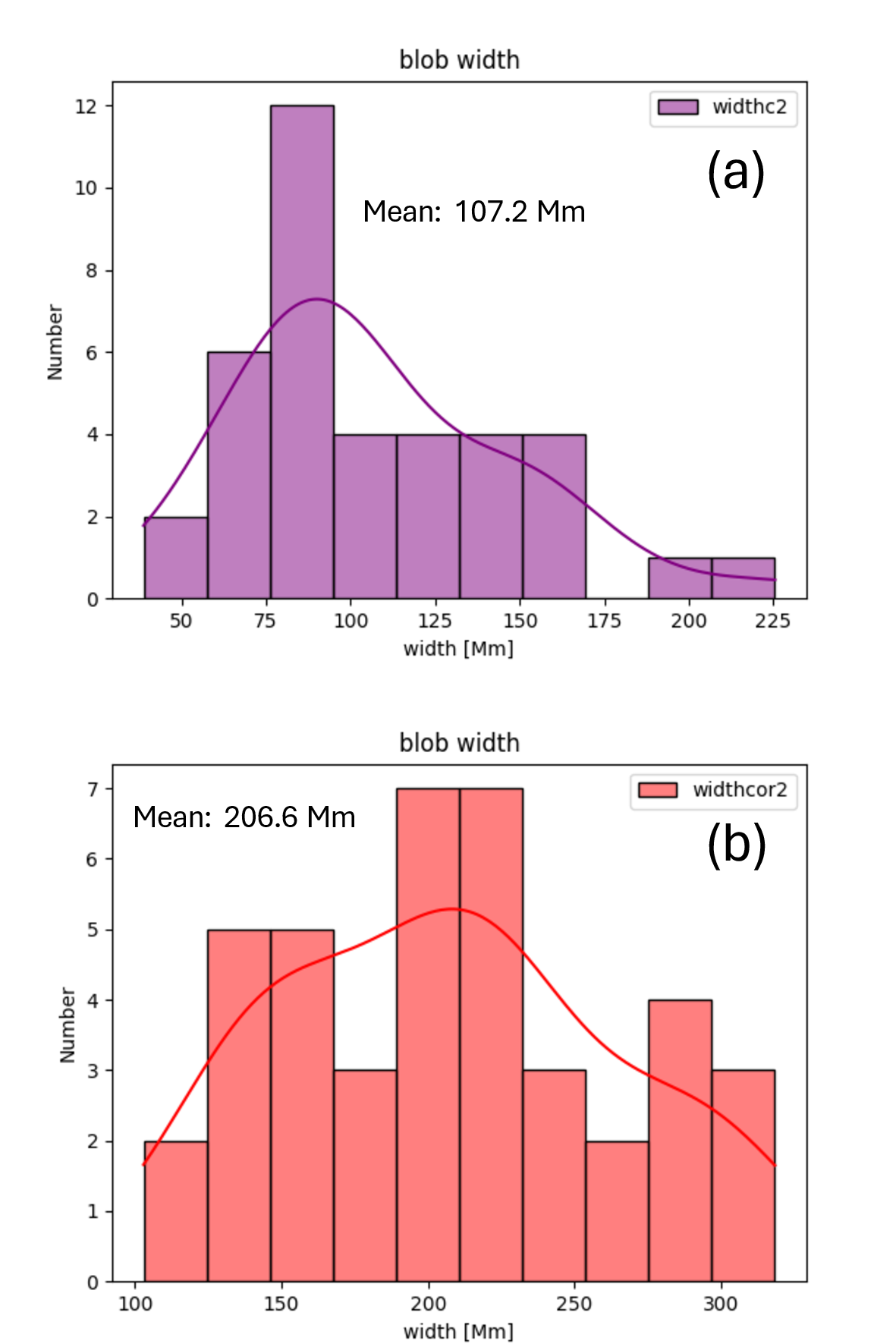}
    \caption{Histograms of blob width in LASCO-C2 (top panel) and COR2 (lower panel) images.
		The corresponding mean values are labeled.}
    \label{fig13}
\end{figure}

For each blob, we derive their widths at every moment.
In Figure~\ref{fig13}, histograms of blob width measured in LASCO-C2 and COR2 images are displayed in the top and bottom panels, respectively.
It is found that the widths range from $\sim$38.8 to $\sim$225.4 Mm with a mean value of $\sim$107.2 Mm in LASCO images.
The widths are between $\sim$102.9 to $\sim$318.3 Mm with a mean value of $\sim$206.6 Mm in COR2 images.
Hence, the maximal width is 3$-$6 times larger than the minimum width.
In Figure~\ref{fig7}(a), the white arrow points to the thin current sheet with a length of $\sim$9\,$R_{\sun}$ above the solar surface.
Three hours later, the current sheet becomes longer with a length of $\sim$12\,$R_{\sun}$ (see panel (b)).
Since the initial formation of current sheet and blobs might be blocked by the occulting disk of WL coronagraphs,
we take the maximal width ($\sim$225.4 Mm) of blobs as the width of current sheet (see Figures~\ref{fig12},~\ref{fig13}).
Consequently, the aspect ratio ($L/\delta$) of current sheet falls in a range of 28$-$37, which is significantly larger than the theoretical threshold (2$\pi$).
The true values of $L/\delta$ may exceed 28 for the following reasons.
Firstly, considering the projection effect, the measured lengths of current sheet in LASCO coronagraphs are less than their true lengths in 3D.
Secondly, both length and width of the current sheet evolve with time. As the CME propagates outward, the current sheet is stretched gradually.
The final length of the current sheet should be larger than that shown in Figure~\ref{fig7}(b).
Thirdly, the spatial resolutions of WL coronagraphs significantly affect the measured widths of current sheets.
Previous numerical simulations have demonstrated that the measured width decreases as the spatial resolution improves.
Therefore, the blobs following the CME are probably generated by tearing-mode instability of the current sheet.
\citet{guo13} analyzed 72 blobs observed by SOHO/LASCO between 2002 January 8 21:32 UT and January 10 10:06 UT.
The widths of these blobs fall in a range of 25$-$135 Mm, whose distribution is close to a Gaussian distribution rather than an exponential function.
\citet{lee20} studied four blobs following a CME observed by LASCO-C2 on 2017 September 10. 
The lateral widths are between $\sim$98 and $\sim$245 Mm with an average value of $\sim$175 Mm, which is close to the average width of blobs in our study.

It is surprising that the mean width of blobs measured in COR2 is $\sim$1.9 times larger than that in LASCO-C2 images.
The inconsistency is probably due to different spatial resolutions of the instruments.
In Table~\ref{tab1}, the pixel sizes of LASCO-C2 and COR2 are $\sim$11$\farcs$60 and $\sim$56$\farcs$65 in this study.
Therefore, more pixels are needed to distinguish blobs in COR2 images. On the other hand, projection effect should be taken into account. 
To illustrate the influence of projection effect on current sheet widths, \citet{lin09} drew a schematic cartoon (see their Fig. 7).
The measured width of a current sheet is minimum when observed edge-on and increases with the angle between the LOS and current sheet axis.
In our study, the current sheet and blobs are probably observed edge-on by LASCO. 
The widths of blobs naturally increase when observed from STA with a separation angle of $\sim$12$\degr$ from the Earth.

One of the physical parameters in magnetic reconnection is the reconnection rate, which is defined as the reconnected magnetic flux per unit time \citep{for00}.
The dimensionless form is the Alfv\'{e}n Mach number $M_{A}=V_{inflow}/V_{A}$, where $V_{inflow}$ signifies the inflow speed into the reconnection region and 
$V_{A}$ denotes the Alfv\'{e}n speed, which is approximately equal to the outflow speed $V_{outflow}$. 
Using the observed $V_{inflow}$ and $V_{outflow}$ in current sheets, the values of $M_{A}$ are reported to be 0.01$-$0.23 \citep{lin05} and 0.055$-$0.20 \citep{tak12}.
Considering that the EUV intensities of inflow are too weak to be detected in most cases, 
a good estimate of $M_{A}$ is the ratio of sheet thickness (width) to length, which are easy to measure in observations \citep{for00}. 
According to this definition, the values of $M_{A}$ are reported to be 0.004$-$0.007 \citep{sea17} and 0.15$-$0.33 \citep{xue18}.
In our study, the speeds of blobs in the current sheet could be obtained readily. However, the inflow speeds are hard to measure in WL images.
Consequently, we utilize the ratio of width to length to calculate the reconnection rate.
As mentioned above, the widths of blobs lie in a range of 38.8$-$225.4 Mm in LASCO images and the length of the current sheet is 9$-$12\,$R_{\sun}$.
Therefore, the value of $M_{A}$ falls in a range of 0.027$-$0.035, which is close to the values in statistical investigations \citep{nar06}.

\section{Discussion} \label{dis}
In the standard flare model, a current sheet forms between the bottom of an eruptive flux rope and the hot post-flare loops \citep{shi95,lin00}. 
The current sheets are characterized by inflows, outflows, localized heating to $\ge$10 MK, significant nonthermal velocities, 
termination shocks and particle accelerations \citep{yoko01,bem08,chen15,ree15,gui16,yan18}.
Occasionally, a train of plasmoids or magnetic islands are created and move along the current sheets as a result of tearing-mode instability \citep{asai04,bar08}.
Magnetic free energy is transferred to small scales via a cascade of consecutively smaller and smaller plasmoids, 
which facilitate particle accelerations efficiently \citep{bar11}.
Meanwhile, small plasmoids may coalesce to form larger and wider plasmoids \citep{oka10,shq12b,tzh26}. 
These blobs should be observed not only in EUV and SXR images \citep{ohy98,tak12,lu22,lsy25}, 
but also in WL images of coronagraphs \citep{ko03,lin05,shq12a,cx18,lee20,bem22}.

In this study, a curved and thin current sheet in the wake of CME is observed on 2023 March 20.
The direction of current sheet is in line with the direction of CME movement.
A total of eight and seven blobs moving outward are identified in LASCO-C2 and STA/COR2 images, respectively. 
The apparent speeds are between 336 and 871 km\,s$^{-1}$, which are in agreement with previous observations \citep{ko03,lin05,shq12a,lee20}.
\citet{bem22} explored the blob observed by Metis coronagraph \citep{ant20} on board Solar Orbiter \citep[SolO;][]{mu20} on 2021 February 13.
The apparent speed of the blob is only $\sim$210 km\,s$^{-1}$, which is much slower than our results.
\citet{ko03} investigated two blobs on 2002 January 9. One of them show considerable acceleration and the other moves at a constant speed.
\citet{lin05} studied 5 blobs on 2003 November 18 in detail. Four of them show acceleration (16$-$158 m\,s$^{-2}$) and only one shows deceleration (-42.5 m\,s$^{-2}$).
\citet{shq12a} analyzed 11 CME events associated with blobs. About 58\% of the blobs experience acceleration, 20\% of them experience deceleration,
and 22\% of them move with nearly constant speeds. Among the four blobs on 2017 September 10, two of them show remarkable acceleration or deceleration \citep{lee20}.
In our study, the blobs show negligible acceleration or deceleration during their propagation.

The maximal heliocentric heights of the blobs range from 2.9 to 7.6\,$R_{\sun}$, which are close to the heights of four blobs on 2017 September 10 \citep{lee20}.
As the blobs move forward, they merge with the background corona and become blurred, which could not be tracked any more.
Unfortunately, there are two LASCO-C3 images during the CME. Hence, it is impossible to track the blobs with LASCO-C3.
\citet{shq12a} measured the heights and lifetimes of blobs after CMEs, most of which reach up to 10$-$28\,$R_{\sun}$ and 20$-$40 hrs.
The lifetimes of the blobs in our study are 7$-$100 minutes, 
which are close to the values reported by \citet{ko03,lin05}, but are significantly shorter than those reported by \citet{shq12a}.
Recently, a spaceborne mission, the Solar Close Observations and Proximity Experiments \citep[SCOPE;][]{lin25}, has been proposed by Chinese scholars.
The spacecraft is designed to get close to the corona at a minimum height of 5\,$R_{\sun}$ with a highly inclined orbit. 
One of the major scientific goals is to detect CME-flare current sheets with unprecedented resolutions. 
Our multipoint observations of current sheet and blobs will give important instructions to the design of the mission.
In the future, multipoint (e.g. L5 point), multiwavelength, and high-resolution observations are required to 
understand current sheets and fine structures within them.

Blobs are ubiquitous in the solar atmosphere, such as in flare current sheets \citep{yan22,cgc24} and coronal jets \citep{man22,ylh24,gu26,ylp26}.
\citet{llp16} studied magnetic reconnection between an eruptive filament and the ambient coronal loops on 2012 January 1.
A thin current sheet hosting several plasmoids is identified. The speeds and widths of the plasmoids are 100$-$300 km\,s$^{-1}$ and 0.58$-$1.23 Mm.
Using ultrafine observations of the Goode Solar Telescope \citep[GST;][]{cao10} in H$\alpha$ red wing,
\citet{cgc24} studied a small-scale flare and recognized two plasmoids with sizes of $\sim$150 km.
\citet{zqm14} found recurrent blobs in coronal jets observed by SDO/AIA. The temperature, average size, and lifetime are $\sim$2.3 MK, $\sim$3 Mm, and 24$-$60 s.
In follow-up works, multiple blobs are discovered along a jet observed by STEREO/EUVI \citep{zqm16} and SDO/AIA \citep{chen22}.
Using high-resolution observations of the Interface Region Imaging Spectrograph \citep[IRIS;][]{dep14} in UV 1330 {\AA}, 
\citet{zqm19} reported even smaller blobs with sizes of 0.33$-$0.98 Mm. These blobs in flare current sheets and jets are believed to 
form as a result of tearing-mode instability in the coronal environment with plasma $\beta \ll 1$ \citep{wyp16,ni17,yang26}.
Therefore, the sizes of blobs cover a wide spectrum from $\sim$0.15 Mm in the chromosphere to $\sim$300 Mm in the corona.
Numerical simulations are required to explore the nature of blobs.

\section{Summary} \label{sum}
In this paper, we report multiwavelength and multipoint observations of a partial filament eruption originating from NOAA AR 13258 on 2023 March 20.
The long, sigmoidal filament is composed of two parts: the eruptive part and remaining part. Both of them are supported by magnetic flux ropes.
The eruption of the eruptive part generates a C4.1 class two-ribbon flare and a wide CME propagating in the northeast direction.
The separation motion of two ribbons persist for $\sim$12 hr 
and the separation speed decreases from 0.26$-$5.62 to 0.13$-$0.53 km\,s$^{-1}$ observed in H$\alpha$ and 304 {\AA}.
A thin, curved current sheet follows the CME and extends in the northeast direction. 
A series of bright and compact blobs flow out along the current sheet, which are simultaneously observed by SOHO/LASCO and STA/COR2 coronagraphs.
In the FOV of LASCO, the apparent speeds, lifetime, maximal height, and width of the blobs are 
378$-$871 km\,s$^{-1}$, 7$-$84 minutes, 2.94$-$6.25\,$R_{\sun}$, and 38.8$-$225.4 Mm, respectively.
In the FOV of COR2, the apparent speeds, lifetime, maximal height, and width of the blobs are
336$-$712 km\,s$^{-1}$, 20$-$100 minutes, 4.09$-$7.61\,$R_{\sun}$, and 102.9$-$318.3 Mm.
A positive correlation is found between the lifetime and maximal height, suggesting that blobs with longer lifetime propagate further from the solar surface.
The total duration of these blobs is $\ge$8 hr, which is accordant with continuous flare reconnection 
manifested by a persistent ribbon separation in the chromosphere.
The aspect ratio (28$-$37) of the current sheet is probably larger than the critical value, 
implying that the blobs following the CME are possibly generated by tearing-mode instability in the current sheet.

\begin{acknowledgements}
The authors appreciate the reviewer for inspiring comments and suggestions to improve the quality of this article.
We also thank Prof. Lei Ni, Prof. Yulei Wang, and Dr. Liping Yang for valuable discussions.
SDO is a mission of NASA\rq{}s Living With a Star Program. AIA and HMI data are courtesy of the NASA/SDO science teams.
STEREO/SECCHI data are provided by a consortium of US, UK, Germany, Belgium, and France.
This CME catalog is generated and maintained at the CDAW Data Center by NASA and 
The Catholic University of America in cooperation with the Naval Research Laboratory. 
SOHO is a project of international cooperation between ESA and NASA.
This work is supported by the Strategic Priority Research Program of the Chinese Academy of Sciences, grant No. XDB0560000,
the National Key R\&D Program of China 2022YFF0503002 (2022YFF0503000, 2022YFF0503001), 2021YFA1600500 (2021YFA1600501, 2021YFA1600502), 
ISSI/ISSI-BJ Team 2024 (24-604), NSFC under the grant numbers 12373065, 12373057, 12403068,
 and Natural Science Foundation of Jiangsu Province (BK20231510, BK20241707).
JC acknowledges the use of data from the Chinese Meridian Project (CMP).
\end{acknowledgements}

\bibliographystyle{raa}
\bibliography{blob}

\label{lastpage}

\end{document}